\documentstyle[12pt]{article}
\def\ptoday{{\sl {\number\day \space de\space \ifcase\month 
\or janeiro\or fevereiro\or mar{\c c}o\or abril\or maio
\or junho\or julho\or agosto\or setembro\or outubro
\or novembro \or dezembro\fi\space de\space \number\year}}}    
\makeatletter
\@addtoreset{equation}{section}
\makeatother

\newcommand{\ddouble}{{\partial^{^{\kern-6pt \leftrightarrow}}}}

\newcommand{\dpad}[2]{{\displaystyle{\frac{\partial #1}{\partial #2}}}}

\newcommand{\dfrac}[2]{{\displaystyle{\frac{#1}{#2}}}}
\newcommand{\dsum}[2]{\displaystyle{\sum_{#1}^{#2}}}
\newcommand{\dint}{\displaystyle{\int}}
\newcommand{\equ}[1]{(\ref{#1})}
\newcommand{\es}{\\[3mm]}
\newcommand{\p}{\psi}
\newcommand{\beq}{\begin{equation}}
\newcommand{\eqn}[1]{\label{#1}\end{equation}}
\newcommand{\ba}{\begin{array}}
\newcommand{\ea}{\end{array}}

\def\a{\alpha}

\def\f{\phi}

\def\m{\mu}

\def\s{\sigma}

\def\x{\xi}

\def\D{\Delta}
\def\F{\Phi}
\def\G{\Gamma}
\def\psi{\Psi}

\def\S{\Sigma}

\def\cb{{\cal B}}

\def\co{{\cal O}}

\def\cs{{\cal S}}

\def\cv{{\cal V}}

\def\cz{{\cal Z}}
\def\inbar{\vrule height1.5ex width.4pt depth0pt}
\def\rlx{\relax\leavevmode}
\def\I{\leavevmode\hbox{\small1\kern-3.8pt\normalsize1}}
\def\openone{\leavevmode\hbox{\small1\kern-3.3pt\normalsize1}}
\def\Ione{\rlx{\rm 1\kern-2.7pt l}}
\def\Ik{\rlx{\rm I\kern-.18em k}}
\def\IC{\rlx\leavevmode
             \ifmmode\mathchoice
                    {\hbox{\kern.33em\inbar\kern-.3em{\rm C}}}
                    {\hbox{\kern.33em\inbar\kern-.3em{\rm C}}}
                    {\hbox{\kern.28em\sinbar\kern-.25em{\rm C}}}
                    {\hbox{\kern.25em\ssinbar\kern-.22em{\rm C}}}
             \else{\hbox{\kern.3em\inbar\kern-.3em{\rm C}}}\fi}
\def\IP{\rlx{\rm I\kern-.18em P}}
\def\IR{\rlx{\rm I\kern-.18em R}}
\def\IN{\rlx{\rm I\kern-.20em N}}

\def\llsymbol#1{\@llsymbol{\@nameuse{c@#1}}}
\def\@llsymbol#1{\ifcase#1\or {}\or {'}\or {''}\or {'''}\or
   {''''}\or {'''''}\or  \else\@ctrerr\fi\relax}
\newcommand{\ol}\overline
\newcommand{\ti}\tilde
\newcommand{\wt}\widetilde
\newcommand{\wh}\widehat
\newcommand{\bv}\breve
\newcommand{\dg}\dagger

\newcommand{\be}{\begin{equation}}
\newcommand{\ee}{\end{equation}}
\newcommand{\bl}{\begin{eqnarray}&}
\newcommand{\el}{&\end{eqnarray}}
\newcommand{\bq}{\begin{eqnarray}}
\newcommand{\eq}{\end{eqnarray}}

\newcommand{\pa}{\partial}

\def\sl#1{\rlap{\hbox{$\mskip 1 mu /$}}#1}
\def\Sl#1{\rlap{\hbox{$\mskip 3 mu /$}}#1}

\begin{document}

{\hfill%
\parbox{50mm}{\large hep-th/9902084\\
                            TUW-99-03\\
                                          }} \vspace{3mm}


\begin{center}
{{\LARGE {\bf Gauge Independence\\[3mm]of the Effective Potential
Revisited}}}

\vspace{7mm}

{\large Oswaldo M. Del Cima$^{\rm (a)}$\footnote{{{Supported by the 
{\it Fonds zur F\"orderung der Wissenschaftlichen Forschung (FWF)} under
the
contract number P11654-PHY.}}}, Daniel H. T. Franco$^{\rm (b)}$\footnote
{{{Supported by the {\it Conselho Nacional de
Desenvolvimento Cient\'\i fico e Tecnol\'ogico (CNPq)}.}}} and 
Olivier Piguet$^{\rm (c)}$\footnote{{{Supported by the 
{\it Conselho Nacional de
Desenvolvimento Cient\'\i fico e Tecnol\'ogico (CNPq)}.}}}}

\vspace{4mm}

$^{{\rm (a)}}$ {\it Institut f\"{u}r Theoretische Physik, \\Technische
Universit\"{a}t Wien (TU-Wien),\\Wiedner Hauptstra{\ss}e 8-10 - A-1040 -
Vienna - Austria.}

$^{{\rm (b)}}$ {\it Scuola Internazionale Superiore di Studi Avanzati
(SISSA),\\Via Beirut 2-4 - 34014 - Trieste - Italy.}

$^{{\rm (c)}}$ {\it Universidade Federal do Esp\'{\i}rito Santo (UFES)
, \\CCE, Departamento de F\'{\i}sica,\\ Campus Universit\'ario
de Goiabeiras - 29060-900 - Vit\'oria - ES - Brasil.}

{\tt E-mails: delcima@tph73.tuwien.ac.at, franco@manon.he.sissa.it, 
piguet@cce.ufes.br}

{PACS numbers: 11.10.Gh, 11.15.-q, 11.15.Bt, 11.15.Ex}

\end{center}


\begin{center}
{\bf Abstract}
\end{center}

{\small  We apply the formalism of {\it extended BRS symmetry} to the
investigation of the gauge dependence of the effective potential in a 
spontaneously symmetry broken gauge theory. This formalism, which 
includes a set of Grassmann parameters defined as the 
BRS variations of the gauge-fixing parameters, allows us to derive
in a quick and unambiguous way the related Nielsen identities, 
which express the physical gauge independence, in a class of 
generalized 't Hooft gauges, of the effective
potential. We show in particular that the validity of
the Nielsen identities does not require any constraint on the gauge-fixing 
parameters, to the contrary of some claims found in the 
literature. We use the method of algebraic renormalization, which leads
to results independent of the particular renormalization scheme used.}

\newpage

\section{Introduction}


The notion of effective potential was first introduced by Euler,
Heisenberg and Schwin\-ger~\cite{euler} and later applied to studies of
spontaneous symmetry breakdown by Goldstone, Salam, Weinberg and
Jona-Lasinio~\cite{goldstone}. Unfortunately, an exact computation of the
effective potential is very hard, often the best answer being given for
the first few terms in a loop expansion~\cite{coleman}-\cite{dolan}. 
This is a difficult task, in particular when several interactions 
are present, as it is the case in spontaneously broken gauge theories.  
In such theories, the calculation of the radiative corrections to the
effective potential has long been of interest, specially in view of its 
gauge dependence~\cite{dolan}-\cite{rama}.

However, the problem of gauge (in)dependence may have been obscured 
by some confusion between the {\it classical gauge invariant potential} 
$\cv_{\rm class}$, used in order to determine, at the classical level, the
field configuration $\f=v$ corresponding to the minimum of the energy 
density, {\it i.e.}, the classical ground state, and the 
{\it effective potential} $\cv_{\rm eff}$, defined after the gauge 
has been ``fixed''. To the contrary of the classical potential, 
the effective potential is gauge dependent, even in the tree 
approximation, {\it i.e.}, it depends on the gauge-fixing parameters 
$\xi$. However its gauge dependence is restricted 
by the following {\it Nielsen identities}~\cite{nielsen}:
\be
\dpad{  \cv_{\rm eff}( \f,\x)}{\xi_\a}+  C_{\a i}( \f,\x) 
\dpad{  \cv_{\rm eff}( \f,\x)}{  \f_i} = 0~~.
\label{nielsen-gen-unshifted}
\ee
Here, the argument $\f$ of $\cv_{\rm eff}$ denotes the set of 
``classical fields'' $\f_i$, $i=1,2,\cdots$, corresponding to
the scalar fields of the theory, and $\x$ the set of gauge-fixing 
parameters $\x_\a$, $\a=1,2,\cdots$. The function $C_{\a i}(\f,\x)$ is 
calculable.

The Nielsen identities imply that the potential
\[
V(\f) = \cv_{\rm eff}(\hat\f(\f,\xi),\xi)~~,
\]
where $\hat\f(\f,\xi)$ is solution of the set of differential equations
\[
\dpad{\hat\f_i}{\xi_\a} = C_{\a i}(\hat\f,\xi)~~,
\]
with some boundary condition $\hat\f(\f,\xi_0)=\f$, is gauge independent:
\[
\dpad{V(\f)}{\xi_\a}  = 0~~.
\]
For suitable boundary conditions, this potential $V(\f)$ coincides, in the
tree approximation, with the classical gauge invariant potential 
$\cv_{\rm class}$.

The problem is particularly of relevance in spontaneously broken gauge
theories quantized with a 't Hooft-like gauge condition implying some
scalar fields. In this case, at the value $ \f=v$ of the 
scalar fields which minimizes the effective potential -- in fact at
any stationary point of the effective potential: 
\be
\left.
\frac{\partial {{ \cv_{\rm eff}}( \phi,\xi)}}{\partial  \phi_i}
\right|_{ \phi=v}=0~~,
\eqn{min-nonshift}
we get, from the Nielsen identities \equ{nielsen-gen-unshifted},
the ``physical'' gauge independence of the effective potential, 
{\it i.e.}, the gauge independence of its minimum:
\be
\left.\dpad{  \cv_{\rm eff}( \f,\x)}{\xi_\a}\right|_{ \phi=v}=0~~.
\eqn{phys-gauge-ind-unshift}

The meaning of the Nielsen identities thus is that
the vacuum that realizes the minimum of the effective potential 
-- {\it e.g.} the spontaneous symmetry breaking -- is a physical minimum.

In this paper, we revisit the problem of the gauge dependence of
the effective potential and extend the discussion to 
all orders of perturbation theory. As we have mentioned in the beginning, 
indeed, in spite of the large number of papers which have looked
carefully at it, a lot of confusion has arisen in the literature,
in the context of a class of gauge models quantized 
with generalized 't Hooft gauges. For instance, it is claimed
in~\cite{fraser} that for a gauge-fixing term of the form
\begin{equation}
\Sigma_{\rm gf}=-\int d^4x\,\,\frac{1}{2\alpha}
\left(\partial^\mu A^{a}_{\mu}+\rho^{ai}\phi_i\right)^2~~,
\end{equation}
where $\rho^{ai}$ and $\a$ are the gauge parameters, the Nielsen 
identities can be derived only if the $\rho^{ai}$'s are
$\a$-independent and if $\rho^{ai}v_i=0$, $v_i$ being the vacuum 
expectation value of $\f_i$. On the other hand, the authors 
of~\cite{johnston,bazeia} have derived the Nielsen identities also 
for the case of $\rho^{ai}$ depending on $\a$, $\rho^{ai}=f(\a)\lambda^{ai}$.
But they still demand that $\lambda^{ai}$ be perpendicular 
to the direction of symmetry breakdown, {\it i.e.}, $\lambda^{ai} v_i=0$. 
In both cases, this condition of transversality arose from the
procedure followed there, which consists in including first 
the gauge-fixing term in the action and only then minimizing the 
potential -- which of course is gauge dependent, already at the 
tree level. Moreover, if one tries to minimize this gauge dependent 
potential, one finds, as presented in~\cite{dolan}, spurious gauge 
dependent solutions corresponding to other stationary points, 
in addition to the usual gauge independent symmetry breaking minima. 
Therefore, for the Nielsen identities to hold -- thus removing these 
unphysical minima -- such constraints as $\rho^{ai}v_i=0$ should be 
imposed, according to the argumentation of~\cite{fraser}-\cite{rama}.
At this point we would like to make another comment concerning 
the orthogonality condition, $\rho^{ai}v_i=0$, imposed, {\it a 
posteriori}, as a necessary condition to have the Nielsen identities 
satisfied by the effective potential. It sounds at least strange such 
a condition by the fact that, if we think in the other way around 
by choosing a particular direction in the space of the gauge 
parameters ($\rho^{ai}$), $\hat\rho^{a\hat i}$, the condition 
$\hat\rho^{a\hat i}\hat{v}_{\hat i}=0$ would establish an 
orthogonal subspace of the ``correct'' directions of 
symmetry breaking, $\hat{v}_{\hat i}$! But the $\hat\rho^{a\hat i}$'s 
are simply gauge parameters, roughly speaking, they could never 
dictate the rules. 

Our aim is to show that, to the contrary of the claims above, 
it is possible to derive the Nielsen identities generally, without any 
restriction on the gauge parameters. In order to achieve this, we define
the vacuum -- characterized by the vacuum expectation value $v$ of 
the scalar field $\f$, around which perturbation theory is developed -- 
at the classical level already, {\it i.e.}, without any gauge dependent 
ambiguity. The gauge-fixing is next introduced in terms of the 
shifted field $\tilde\f=\f-v$, which has a vanishing vacuum 
expectation value.

We shall derive in this way the Nielsen identities for a general 
non-Abelian Yang-Mills gauge theory with scalar and 
spinor matter fields in a 
't Hooft-like gauge. We use the techniques of ``extended BRS 
invariance''~\cite{zuber,piguet-sibold,pigsor}, which
has been introduced precisely in view of investigating and controlling
the gauge dependence in any gauge theory, in particular 
the gauge independence of the physical quantities.
As we shall see, the Nielsen identities follow straightforwardly from
the Slavnov-Taylor identity associated to extended BRS 
invariance\footnote{The formalism of extended 
BRS symmetry has also been used by the author of~\cite{johnston}.}. 
Although we specialize the analysis to the case of a semi-simple Lie group 
for the sake of simplicity, our results obviously hold for the 
case of a general compact gauge group too, provided all the fields 
remain massive. The renormalizability of such a theory has indeed 
been proven in~\cite{bbbc}, the generalization to extended 
BRS invariance being straightforward. 

\noindent {\bf Note.} 
A derivation of the Nielsen identities, based on BRS invariance, has 
already been given in~\cite{das}, for the Abelian Higgs model in the 
't Hooft gauge, and in~\cite{metaxas} for a more general gauge
model. Our presentation, however, differs from the latters in the amount 
that we emphasize the use of extended BRS invariance, which allows to 
derive the result in a quick and elegant way and staying on the 
firm basis of rigorous renormalization theory.

The plan of the paper is as follows. First, in order to provide the
necessary basis for the unfamiliarized  reader, we review 
in Section 2 the construction of the BRS and extended BRS invariant 
theory in the tree-graph approximation, described by a classical action 
obeying functional identities characterizing the gauge-fixing and the 
(extended) BRS invariance. We show, at the end of this section, how 
extended BRS invariance does control the gauge (in)dependence. The 
renormalized theory is described at the beginning of Section 3, which 
continues with the derivation of the Nielsen identities for the 
effective potential.

\noindent {\bf N.B.} In order to avoid infrared problems in the
definition of the effective potential, we shall work with massive fields
only. The  mechanism of spontaneous symmetry breaking 
will thus be assumed to
provide nonzero masses to all physical fields -- scalars, fermions and 
gauge bosons. The choice of a generalized 't Hooft gauge then ensures 
nonvanishing masses for all unphysical (ghost) fields which, in any case, 
decouple from the physical sector of the theory due to BRS invariance,
as it is well-known~\cite{kugo-ojima,becchi-houches}.


\section{Extended BRS Symmetry in the Classical Approximation}
\label{sect2}
\subsection{The Classical Theory: Spontaneous Breakdown 
of Gauge Invariance}

Matter is described by a set of scalar and spinor fields, 
$\phi_i(x)$ ($i=1,\dots,n$) and $\psi_I(x)$ ($I=1,\dots,N$), respectively, 
belonging to some unitary representation of a semi-simple Lie group 
$G$\footnote{It should be stressed that, as pointed out previously, it 
could be considered here a nonsemi-simple gauge group, but for the
sake of simplicity we restrict our case to a semi-simple one.}. 
The matter fields carrying an anti-Hermitian fully reducible 
representation of $G$ transforms as
\bq
&&\delta \phi_i(x) =\omega^a(x)
T_a^{(\phi)}{}_i{}^j\phi_j(x)~~,\label{pg1a}\\
&&\delta \psi_I(x) =\omega^a(x)
T_a^{(\Psi)}{}_I{}^J\psi_J(x)~~,~~~\delta \bar\psi_I(x) =-\omega^a(x)
T_a^{(\Psi)}{}_I{}^J\bar\psi_J(x)~~,  \label{pg1}
\eq
where the matrices $T_a$ are anti-Hermitian and obey the commutation
relations of the Lie algebra of the group $G$: 
\begin{equation}
\left[ T_a,\,T_b\right] =\,f_{ab}^{\;\;c}~T_c~~.  \label{pg2}
\end{equation}

Because of the local character of the transformations one has to introduce
covariant derivatives 
\bq
D_\mu\phi_i(x)=\partial_\mu \phi_i(x)-A_\mu^a(x)
T_a^{(\phi)}{}_i{}^j\phi_j(x)~~,\\
{\Sl D}\psi_I(x)=\sl\partial \psi_I(x)-{\Sl A}^a(x)
T_a^{(\Psi)}{}_I{}^J\psi_J(x)~~,  
\label{pg3}
\eq
the connection being given in terms of vector fields $A_\mu^a(x)$
transforming as 
\begin{equation}
\delta A_\mu^a(x)=\partial_\mu \omega^a(x)-f_{\,\,\,bc}^{\,a\;\;}
\omega^b(x)A_\mu^c(x)~~.
\label{pg4}
\end{equation}

A gauge invariant action is built up with the matter fields, $\phi_i$ and 
$\psi_I$, and the gauge fields, $A_\mu^a$, as 
\bq
{\Sigma}_{{\rm inv}}=\int d^4x
~~\biggl\{\!\!\!\!\!\!\!\!\!\!&&-\frac 1{4g^2}F_{\mu\nu}^aF_a^{\mu\nu} 
+i\bar{\Psi}_I{\Sl D} \psi_I+m_{IJ}\bar{\psi}_I\psi_J
+D^\mu \phi_i D_\mu \phi_i-\mu_{ij}^2 \phi_i \phi_j + \nonumber \\ 
&&+\lambda_{IJk}\bar{\psi}_I\psi_J\phi_k
-h_{ijkl}\phi_i\phi_j\phi_k\phi_l\biggr\}~~,  
\label{pg5}
\eq
where $F_{\mu\nu}^a=\partial_\mu A_\nu^a-\partial_\nu A_\mu^a-
f_{\,\,\,bc}^{\,a\;\;}A_\mu^bA_\nu^c$ and the Yukawa and quartic 
coupling constants, $\lambda_{IJk}$ and $h_{ijkl}$, respectively, 
are invariant tensors of the gauge group $G$.

The form of the potential, 
\be
{\cal V}\left( \phi \right) = \mu_{ij}^2 \phi_i \phi_j +
h_{ijkl}\phi_i\phi_j\phi_k\phi_l~~,
\eqn{class-pot}
is chosen such as to ensure the broken regime, {\it i.e.}
$\mu _{ij}^2<0$, and $h_{ijkl}$ must be positive-definite 
for the sake of stability of the system, guaranteeing, therefore, 
the validity of perturbation theory.
In the classical theory, the potential is the energy density 
for constant scalar fields  -- all other fields vanishing. The
equilibrium state (the ``fundamental state'', or ``vacuum state'') is
given by the field configuration which minimizes the energy density. 
This minimum is obtained at some value\footnote{In fact, the 
condition of minimum does not determine $v_i$ uniquely, but only up to
a group transformation: it fixes an orbit in the space of 
scalar fields. One has to choose one particular -- arbitrary -- value 
in the orbit.} $\f_i=v_i$ function of the parameters 
$\mu_{ij}^2$, $h_{ijkl}$. This value is interpreted in the 
corresponding quantum theory as the vacuum expectation value of the 
field $\phi$. The equilibrium -- or vacuum --
state not being invariant under the gauge transformations, gauge 
invariance is said to be spontaneously broken.

In order to study the small oscillations around the equilibrium/vacuum
state -- which gives rise to the physical particle interpretation 
in the quantum case -- one proceeds to the change of variables 
\begin{equation}
\phi_i=v_i+\tilde{\phi}_i~~,  \label{pg10}
\end{equation}
where $\tilde{\phi}_i$ are the Higgs scalars, all with 
vanishing vacuum expectation values, such that
\begin{equation}
\left.
\frac{\partial \tilde\cv_{\rm eff}(\tilde\f)}
{\partial \tilde{\phi _i}}\right|_{\tilde{\phi}=0}=0~~,\quad
\mbox{where}\quad 
\tilde\cv_{\rm eff}(\tilde\f) = {\cal V}(v+\tilde{\phi})~~.  
\label{pg10a}\end{equation}
Let us recall that the mass matrix of the scalar fields satisfies the 
eigenvalue equation 
\begin{equation}
M_{ij}^2\left( T^a\right) _{\,\,k}^jv^k=
 \left. \frac{\partial^2{\tilde\cv_{\rm eff}}
(\tilde{\phi}) }{\partial \tilde{\phi_i}
\partial \tilde{\phi}_j}\right|_
{\tilde{\phi}=0}\left( T^a\right) _{\,\,k}^jv^k=0~~,  \label{pg10b}
\end{equation}
whereas the mass matrix $m_{ab}^2$ of the gauge vector fields is given by
\begin{equation}
m_{ab}^2=\left( T_a\right) _{\,\,j}^iv^j\left( T_b\right)
_{\,\,k}^iv^k~~.
\label{pg10c}
\end{equation}
In the following we shall assume that all the vector fields 
acquire a mass. For the scalar fields the same will be achieved 
by a suitable choice of the gauge-fixing condition.

\subsection{Gauge-fixing}

Since from now on we shall work in terms of the shifted fields 
$\tilde\f_i$ (the Higgs scalars) with vanishing vacuum 
expectation value, we will omit, therefore, the tilde symbol.

In order to quantize the theory one has to fix the gauge. We first
require invariance under the following BRS transformations:
\begin{equation}
\begin{array}{ll}
sA_\mu ^a=D_\mu c^a\equiv \left( \partial _\mu
c^a-f_{\,\,\,bc}^{\,a\;\;}A_\mu ^bc^c\right) \,,\qquad & sc^a=\frac
12f_{bc}{}^ac^bc^c\,, \\[3mm] 
s\Psi_{I}=\,c^aT_a^{(\Psi)}{}_I{}^J\Psi_J\,\,,\;\;\;s\bar{\Psi}_{I}=\,
\bar{\Psi}_JT_a^{(\Psi)}{}_I{}^Jc^a\,\,,\qquad & s{\bar{c}}^a=b^a\,, 
\\[3mm] 
s{\phi}_i=\,c^aT_a^{({\phi})}{}_i{}^j(v_j+{\phi}_j)
\,\,,\;\;\;\;\;\;\;\;\;\qquad & sb^a=0~~.
\end{array}
\label{pg11}
\end{equation}
The BRS transformations of the matter and gauge fields are their 
gauge transformations \equ{pg1a}, \equ{pg1} and \equ{pg4}, with 
the infinitesimal parameters $\omega^a(x)$ being replaced by the 
anticommuting Faddeev-Popov ghost fields $c^a(x)$. We have also
introduced the antighost fields, ${\bar{c}}^a(x)$, and the Lagrange
multiplier fields, $b^a(x)$, which will be used in order to 
define the gauge-fixing condition. The transformation of the 
ghosts, $c^a$, was chosen such as to make the BRS operator nilpotent:
\[
s^2=0~~. 
\]
The gauge-fixing is then defined through the introduction in the action of
the gauge breaking term -- BRS invariant due to the nilpotency
of $s$:
\begin{eqnarray}
\Sigma_{{\rm gf}} &=&\,{\int }d^4x\,\,\left\{ b_a\left( \partial ^\mu
A_\mu^a+\rho ^{ai}{\phi}_i\right) +\frac 12\alpha b_ab^a-\bar{c}_a\left[
\delta _b^a\partial ^\mu D_\mu +\rho
^{ai}T_b^{({\phi})}{}_i{}^j(v_j+%
{\phi}_j)\right] c^b\right\} \,  \nonumber \es
&=&s\,{\int }d^4x\,\,\left\{ \bar{c}_a\left( \partial ^\mu A_\mu ^a+\rho
^{ai}%
{\phi}_i\right) +\frac 12\alpha \bar{c}_ab^a\right\}~~,
\label{pg12}\end{eqnarray}
where $\rho _{ai}$ and $\alpha $ are the ``gauge parameters''. 
For a  generic value of the 't Hooft parameters $\rho _{ai}$, 
all scalar fields become massive. 

\subsection{Gauge Independence and Extended BRS Invariance}

One observes that the gauge dependence of the classical theory
is given by
\be
\frac{\partial \Sigma} {\partial \alpha } = s\int d^4x\,\,
\frac12\bar{c}_ab^a~~,  \qquad
\frac{\partial \Sigma} {\partial \rho ^{ai}} 
 =s\int d^4x\,\, \bar{c}_a{\phi}_{i\,\,}~~,
\eqn{pg13}
where $\S$ is the total action, sum of \equ{pg5} and \equ{pg12}.
The right-hand sides of (\ref{pg13}) appear as a BRS-variation, which 
expresses the unphysical character of the gauge parameters. This means
that the physical quantities such as the $S$-matrix elements and the Green 
functions of gauge invariant operators are independent of these 
parameters~\cite{pigsor}.

In order to translate later on these equations into a functional form, 
we introduce new Grassmann variables, $\chi$ and $\eta_{ai}$,
and define the {\it BRS transformations of the gauge parameters} 
as follows:
\be\ba{l}
s\alpha =\chi \,\,,\;\;\;s\chi =0\,\,,  \es
s\rho _{ai} = \eta _{ai}\,\,,\;\;\;s\eta _{ai}=0\,\,.  
\ea\eqn{pg14}
We shall now require invariance under the 
``extended BRS transformations''~\cite{zuber,piguet-sibold,pigsor}, 
{\it i.e.}, under the transformations \equ{pg14} taken 
together with the field BRS transformations \equ{pg11}. 
This implies the modification of the gauge breaking term \equ{pg12} into:
\begin{eqnarray}
\Sigma_{{\rm gf}} 
&=& s \,{\int }d^4x\,\,\left\{\bar{c}_a\left(\partial^\mu A_\mu^a
+\rho^{ai}{\phi}_i\right) +\frac 12\alpha \bar{c}_ab^a\right\} \nonumber\es
&=&\,{\int }d^4x\,\,\left\{b_a\left( \partial ^\mu
A_\mu^a+\rho ^{ai}{\phi}_i\right) +\frac 12\alpha b_ab^a
-\bar{c}_a\left[\delta _b^a\partial ^\mu D_\mu +\rho
^{ai}T_b^{({\phi})}{}_i{}^j(v_j+%
{\phi}_j)\right] c^b+\right.  \nonumber \es
&&\left. +\,\frac 12\chi \bar{c}_ab^a+\eta ^{ai}\bar{c}_a{\phi}%
_{j\,\,}\right\}~~.  \label{pg151}
\end{eqnarray}
The {\it extended BRS invariance} will allow us to control the 
gauge parameter dependence of the theory, in particular it will 
automatically ensures the conditions \equ{pg13}, as we shall show 
in Subsection \ref{gauge-ind}.
\subsection{The Functional Identities}\label{funct-ident}

The BRS symmetry\footnote{From now on ``BRS'' will 
mean ``extended BRS''.} of the model, as well as the gauge-fixing 
we have chosen may be expressed as functional identities obeyed 
by the classical action \equ{pg15} defined below.

Let us first write down the Slavnov-Taylor identity expressing the 
BRS invariance of the theory. Because of the nonlinearity of some of 
the BRS transformations (\ref{pg11}), we have to add to the action a 
term giving their couplings with external fields, the ``antifields'',  
$A_a^{*\mu }$, $c_a^{*}$, $\Psi_I^{*}$, ${\phi}_i^{*}$: 
\begin{equation}
\Sigma_{{\rm ext}}={\int }d^4x{\sum_{\Phi =A_\mu^a,\,c^a,\,\Psi_I,\,
\phi_i}}\Phi^{*}s\Phi ~~.  \label{pg141}
\end{equation}
The antifields are BRS invariant. Thus, from now on, the total classical 
action is given by 
\begin{equation}
\Gamma^{(0)} =\Sigma _{{\rm inv}}+\Sigma _{{\rm gf}}+
  \Sigma _{{\rm ext}}~~,
\label{pg15}
\end{equation}
such that its BRS invariance is expressed through the 
Slavnov-Taylor (ST) identity~\cite{piguet-sibold} 
\begin{equation}
{\cal S}(\Gamma^{(0)})\,\,=\,\, \dint d^4x 
\left( \sum_{\Phi =A_\mu^a,\,c^a,\,\Psi_I,\,\phi_i}
\dfrac{\delta \Gamma^{(0)}}{\delta\Phi^*}
\dfrac{\delta \Gamma^{(0)}}{\delta \Phi } 
+ b^a\dfrac{\delta\Gamma^{(0)}}{\delta \bar{c}{}^a} \right)
+\chi 
\frac{\partial \Gamma^{(0)}}{\partial \alpha }+\eta ^{ai} \frac{\partial
\Gamma^{(0)}}{\partial \rho ^{ai}}=0~~,  
\label{pg16}
\end{equation}
For later use we introduce the linearized ST operator defined as
the derivation of the nonlinear operator $\cs$, 
\[
{\cal B}_{\Gamma ^{(0)}}=\frac{\partial {\cal S}(\Gamma ^{(0)})}{\partial
\Gamma ^{(0)}}~~,  
\]
{\it i.e.}:
\be
{\cal B}_{\Gamma ^{(0)}}\,\,=\,\,\int d^4x\,\,\left\{ 
\sum_{\Phi =A_\mu^a,\,c^a,\,\Psi_I,\,\phi_i}
\left({\ \ \dfrac{\delta \Gamma ^{(0)}}{\delta \Phi ^{*}}}
{{\frac \delta {\delta \Phi }}}+{{\frac{\delta \Gamma^{(0)}}
{\delta \Phi }}}{\frac \delta {\delta \Phi ^{*}}} \right) 
+ b^a\dfrac{\delta}{\delta \bar{c}{}^a} \right\} 
+ \chi \frac \partial {\partial \alpha }
+\eta ^{ai}\frac \partial {\partial \rho^{ai}}~~. 
\label{pg17}
\ee
The operators ${\cal S}$ and ${\cal B}_\Sigma$ obey the 
algebraic identities
\begin{equation}
{\cal B}_{{\cal F}}\,{\cal S}({\cal F})=0\;\;,\;\;\;\forall \;{\cal
F}\;\;\;,
\label{pg19}
\end{equation}
\begin{equation}
\left( {\cal B}_{{\cal F}}\right) ^2=0\;\;\;{\mbox{if}}\;\;\;{\cal
S}({\cal F})=0~~.  
\label{pg20}
\end{equation}
In particular, since the action $\Gamma ^{(0)}$ obeys the  ST\
identity (\ref{pg16}), we have the nilpotency property (\ref{pg20}): 
\begin{equation}
\left( {\cal B}_{\Gamma ^{(0)}}\right) ^2=0~~.  \label{pg21}
\end{equation}

In addition to the ST identity (\ref{pg16}), the action (\ref
{pg15}) satisfies the following constraints:

\noindent -- the gauge condition: 
\begin{equation}
{{\frac{\delta \Gamma^{(0)}}{\delta b_a}}}
=\partial^\mu A_\mu^a+\rho^{ai} {\phi}_i+\alpha \,b^a
+\frac 12\chi \,\bar{c}^a~~,  \label{pg22}
\end{equation}

\noindent -- the ghost equation, which follows from the 
former by commuting the functional derivation 
$\delta/\delta b_a$ with the ST identity
(\ref{pg16}): 
\begin{equation}
{\cal G}_a\Gamma ^{(0)}=\left({{\frac \delta {\delta
\bar{c}_a}}}+\partial^\mu {\frac \delta {\delta A_a^{*\mu }}}
+\rho^{ai}\frac \delta{\delta {\phi}_i^{*}}\right)\Gamma^{(0)}
=-\frac 12\chi \,b^a-\eta ^{ai}{\phi}_i~~.  \label{pg23}
\end{equation}
It is worth noting that, the right-hand sides of
eqs.(\ref{pg22})--(\ref{pg23}) being linear in the quantum fields, 
will not get renormalized.

Notice that the classical action obeys the decomposition 
\begin{equation}
\Gamma^{(0)}=\widehat{\Gamma}^{(0)}+\int d^4x\,\,
\left\{b_a\left( \partial^\mu
A_\mu ^a+\rho ^{ai}{\phi}_i\right) +\frac 12\alpha b_ab^a+\frac
12\chi 
\bar{c}_ab^a+\eta ^{ai}\bar{c}_a{\phi}_{i\,\,}\right\}~~,
\end{equation}
where $\widehat{\Gamma}^{(0)}$ satisfies the homogeneous gauge condition
and the homogeneous ghost equation 
\begin{equation}
{{\frac{\delta \widehat\Gamma^{(0)}}{\delta b_a}}}=0 \ ,
\quad {\cal G}_a\widehat{\Gamma}^{(0)}=0~~,
\end{equation}
which means that $\widehat{\Gamma}^{(0)}$ is independent from $b_a$ and
depends on $\bar{c}_a$ and on the antifields, $A_a^{*\mu}$ and $\phi^*_i$, 
only through the combinations 
\begin{equation}
{\hat{A}}_a^{*\mu}=A_a^{*\mu}+\partial^\mu \bar{c}_a\,,\;\;\;
\phi_i^{*}={\phi}_i^{*}-\rho_{ai}\bar{c}^a~~.  \label{combination1}
\end{equation}

\subsection{Gauge Independence}\label{gauge-ind}

Gauge independence follows from differentiating the ST identity
(\ref{pg16}) with respect to $\chi$ and $\eta^{ai}$ and later setting 
$\chi=\eta^{ai}=0$:
\begin{equation}
\left. \frac{\partial \Gamma ^{(0)}}{\partial \alpha }\right| _{\chi =\eta
=0}=\left. {\cal B}_{\Gamma ^{(0)}}\left( \frac{\partial \Gamma ^{(0)}}{%
\partial \chi }\right) \right| _{\chi =\eta =0}\,\,,\;\;\;\left. \frac{%
\partial \Gamma ^{(0)}}{\partial \rho ^{ai}}\right| _{\chi =\eta
=0}=\left. 
{\cal B}_{\Gamma ^{(0)}}\left( \frac{\partial \Gamma ^{(0)}}{\partial \eta
^{ai}}\right) \right| _{\chi =\eta =0}~~.  \label{pg26}
\end{equation}
Thus, the dependence of the theory on $\chi$ and $\eta^{ai}$ is
automatically restricted to a BRS-variation, as already announced, 
which means that the physical quantities do not depend on the gauge 
parameters.
In other words, extended BRS invariance takes care of the nonphysical 
character of the gauge parameters, and the problem of the gauge 
independence is reduced to the problem of implementing the  ST 
identity to all orders. 

In order to get a physical interpretation, let us write the 
ST identity in terms of the generating functional of the Green 
functions\footnote{In the classical approximation considered 
in this section, the Green functions are made of tree-graph 
contributions only.}$^,$\footnote{The generating 
functional of the connected Green functions is obtained from the 
generating functional of the vertex functions -- coinciding with the
classical action in the tree-graph approximation -- through a Legendre
transformation with respect to the dynamical fields $\F$. Exponentiation
then yields the generating functional of the general Green functions.
See, {\it e.g.}, \cite{itz-zu,pigsor}.}, 
$Z(J_\F,\F^*,q_I)$, where $J_\F$ denotes the sources of the fields $\F$,
$\F^*$ the associated antifields and $q_I$ a set of BRS invariant 
sources coupled to the gauge invariant operators\footnote{A gauge 
invariant operator is defined as an equivalence class of BRS-invariant 
operators modulo BRS-variations -- what is called a cohomology class 
of the nilpotent BRS operator.} of the theory. 
The ST identity \equ{pg16} then writes
\begin{equation}
{\cal S} Z(J_\F,\F^*,q_I)\,\,=\,\, \dint d^4x 
\left( \sum_{\Phi =A_\mu^a,\,c^a,\,\Psi_I,\,\phi_i}
J_\F \dfrac{\delta Z}{\delta\Phi^*}
+J_{\bar{c}{}^a} \dfrac{\delta Z}{\delta J_{b^a}} \right)
+\chi \frac{\partial Z}{\partial \alpha }
+\eta ^{ai} \frac{\partial Z}{\partial \rho ^{ai}}=0~~. 
\label{slavnov-z}\end{equation}
In case of vanishing gauge invariant sources $q_I$, one may deduce from 
\equ{slavnov-z} the gauge independence of the $S$-matrix:
\be
\dpad{}{\a} S = \dpad{}{\rho^{ai}} S = 0~~,
\eqn{g-ind-S}
whereas for vanishing sources, $J_\F$ and $J_{\bar{c}{}^a}$, one gets the 
gauge independence of the generating functional of the Green functions 
of the gauge operators:
\be
\dpad{}{\a} Z(0,0,q_I) = \dpad{}{\rho ^{ai}} Z(0,0,q_I) = 0~~.
\eqn{g-ind-Z}

\section{The Nielsen Identities}
\subsection{Renormalization}
For the classical -- or tree-graph -- approximation of the theory 
described previously, the renormalization program consists in preserving 
all the symmetry properties of the classical theory in the 
perturbative construction of a quantum theory. It is well 
known\footnote{See~\cite{pigsor} and the references to 
the original literature therein.} that this is feasible for the
class of models considered in the present paper, up to a possible
obstruction by the Adler-Bardeen gauge anomaly -- which we shall suppose
to be absent\footnote{This amounts to choose a convenient representation
for the spinor fields~\cite{pigsor}.}.

Concretely, the resulting renormalized theory is given by the vertex 
functional or generating functional of amputated 1-particle irreducible 
Green functions 
\be
\G(A,\p,\f,c,\bar c,b,A^*,\p^*,\f^*,c^*) = 
\G^{(0)}(A,\p,\f,c,\bar c,b,A^*,\p^*,\f^*,c^*) + \co(\hbar)~~,
\eqn{gamma}
which, in the limit $\hbar=0$, coincides with 
the classical action \equ{pg15} and corresponds to the tree-graph
approximation\footnote{Perturbation theory as usual is ordered according
to the number of loops in the Feynman graphs or, equivalently, to the
powers of $\hbar$.}.
The vertex functional $\G$ obeys all the functional identities 
depicted in Subsection \ref{funct-ident} and which define the theory, 
namely, the ST identity \equ{pg16}, the gauge condition \equ{pg22} 
and the ghost equation \equ{pg23}.

\subsection{The Effective Potential and the Nielsen Identities}
The control of the gauge dependence of the Green functions
is given by the identities \equ{pg26} for $\G$ which, as we have 
already mentioned, follow from differentiating the ST identity 
\equ{pg16} with respect to the gauge parameters:
\begin{equation}
\frac{\partial \Gamma}{\partial \xi} =
  {\cal B}_{\Gamma} \left( \frac{\partial \Gamma}{\partial \s }\right)
~~,~~~ \xi = \a,\rho^{ai}~~,~~~\s=\chi,\eta^{ai}~~.
\label{g-indep}
\end{equation}
By setting from now on
\[
\chi=\eta^{ai}=0~~,
\]
we can write eq.\equ{g-indep} explicitly as (see eq.\equ{pg17} 
for the definition of the linearized ST operator $\cb_\G$)
\be
\frac{\pa\G}{\pa\xi} = 
 \dint d^4x \left( \dsum{\Phi =A_\mu^a,\,c^a,\,\Psi_I,\,\phi_i}{}
\ \ \dfrac{\delta \Gamma }{\delta \Phi ^{*}}
         \dfrac{\delta}{\delta \Phi }
  +\dfrac{\delta \Gamma }{\delta \Phi }
         \dfrac{\delta}{\delta \Phi^{*}} 
+b^a \dfrac{\delta}{\delta \bar{c}^a}\right) 
 \Delta_\xi \cdot \Gamma ~~,~~~ \xi = \a,\rho^{ai} ~~,
\eqn{pg31a}
where the operator insertion in the right-hand side, 
$\Delta_\xi\cdot\Gamma$, is defined by
\be
\left.\dpad{\G}{\s}\right|_{\s=0} 
= \D_\xi\cdot\G ~~,\quad \s=\chi,\eta^{ai}~~.
\eqn{op-insertion}

The definition of the effective potential~\cite{poteff,itz-zu} 
involves the vertex functional $\G$, with the 
dependence of $\G$ restricted only to the scalar fields $\f_i$ 
as follows\footnote{Recall that the 
arguments of the vertex functional $\G$, the ``classical fields'', are 
Schwartz fast decreasing test functions.}:
\be\ba{l}
\G_{\rm scalar}(\f) = 
\left.\G\right|_{A_\m=\p=c=\bar c=b=0,\,{\rm and\ all\ }\F^*=0}\es
= \dsum{n=0}{\infty}~\dsum{i_1,\cdots,i_n}{}
 \dfrac 1{i_1!\cdots i_n!}\dint d^4x_1\cdots d^4x_n\,\,
  \Gamma^{i_1\cdots i_n} ( x_1,\cdots,x_n) 
   \phi_{i_1}(x_1) \cdots \phi_{i_n}(x_n)~~.  
\ea\eqn{pg27}
One defines the effective potential as the zeroth order term 
$\cv_{\rm eff}(\f)$ in the expansion of $\G_{\rm scalar}(\f)$ 
involving higher and higher derivatives in the fields $\f_i$:
\begin{equation}
\Gamma_{\rm scalar}(\phi) =\int d^4x\,\,\left\{ 
- \cv_{\rm eff}(\f) + \cz^{ij}(\phi) \partial_\mu\phi_i\partial^\mu\phi_j
    +\cdots \right\}~~,
\label{pg28}
\end{equation}
where the first term involves the sum of all proper functions 
at zero external momenta, the second sums all second derivatives 
at the same point, and so on. In principle the functions $\f_i(x)$ 
remains arbitrary. However, since we wish to compute $\cv_{\rm eff}$, 
this can be achieved by assuming $\f_i(x)$ constant, $\f_i(x)=\f_i$. 
Now, bearing this in mind, the effective potential can be written as
\be
\cv_{\rm eff}(\f) = -\dsum{n=0}{\infty}~\dsum{i_1,\cdots,i_n}{}
 \dfrac 1{i_1!\cdots i_n!}
  {\tilde\Gamma}^{i_1\cdots i_n} (0,\cdots,0)
    \phi_{i_1}\cdots \phi_{i_n}~~,
\eqn{pot-eff-expansion}
where the ${\tilde\Gamma}^{i_1\cdots i_n} (0,\cdots,0)$'s are the 
momentum-space vertex functions taken at zero external momenta.

It then follows from the definition of the effective potential
and from the gauge dependence equations \equ{pg31a} for the 
vertex functional, that the gauge dependence of the effective 
potential $\cv_{\rm eff}(\f)$ is given by
\begin{equation}
\frac{\partial {\cal V}_{\rm eff}}
{\partial \alpha }+C_i(\phi,\alpha) 
\frac{\partial {\cal V}_{\rm eff}}{\partial \phi_i }{\ }
=0~~~,~~~~\frac{\partial {\cal V}_{\rm eff}}
{\partial \rho ^{ai}}+C_{aij}(\phi,\rho) 
\frac{\partial {\cal V}_{\rm eff}}{\partial \phi_j }{\ }=
0~~~,  \label{nielsen}
\end{equation}
where 
\[
C_i(\phi,\alpha)=-\left.\int d^4x~\frac{\delta(\Delta_\alpha
\cdot \Gamma)}{\delta\phi_i^{*}}\right|_{\phi_i(x)=\phi_i}~~~,~~~~
C_{aij}(\phi,\rho)=-\left.\int d^4x~
\frac{\delta(\Delta_\rho\cdot\Gamma)_{ai}}{\delta\phi_j^{*}}
\right|_{\phi_i(x)=\phi_i}~~~. 
\]
The equations (\ref{nielsen}) are the Nielsen 
identities~\cite{nielsen} announced in the Introduction. As we 
have previously mentioned, there are no constraints on the 
space of gauge parameters to be imposed in order to have the 
Nielsen identities satisfied by the effective potential. This 
could be suspected by the main purpose of Nielsen identities, 
the control of the gauge dependence of the effective potential.

\section{Conclusions}
We have thus be able to show how simply and unambiguously 
the application of the idea of extended BRS 
invariance~\cite{zuber,piguet-sibold} to the study of the 
gauge dependence of the effective potential leads to the Nielsen
identities~\cite{nielsen} which control this dependence. 
In particular, for theories quantized in a generalized 't Hooft 
gauge, if one properly defines the theory, no restriction on 
the gauge parameters is required at all, contrary to some claims 
already published in the literature~\cite{fraser}-\cite{rama}.
This algebraic proof of the validity of the Nielsen identities 
at all orders in perturbation theory is independent of any 
particular renormalization scheme. In spite of this proof was 
done in the context of simple gauge groups, the generalization 
to general compact gauge groups is straightforward, since the 
renormalizability of such theories has been rigorously shown 
in~\cite{bbbc}.
 
\subsection*{Acknowledgements}

The authors are very grateful for an invitation to participate to the
Erwin Schr\"o\-din\-ger Institute (ESI) Workshop, 
{\it Quantization, generalized BRS cohomology and anomalies}, 
held in Vienna, October 1998, where this work has been initiated.
One of us (O.P.) thanks a grant from the {\it FWF} 
(contract number P11654-PHY) for his traveling expenses. 
(O.P.) and (D.H.T.F.) thank the ESI for their living expenses. 
(D.H.T.F.) wishes to thank Prof. L. Bonora for his kind
invitation at the Scuola Internazionale Superiore di Studi Avanzati 
(SISSA) in Trieste, Italy. (O.M.D.C.) dedicates this work 
to his wife, Zilda Cristina, to his daughter, Vittoria, and 
to his son, Enzo. He also dedicates it to his mother, Victoria.
The authors would like to thank J.A. Helay\"{e}l-Neto and M. 
Schweda for discussions. They are deeply indebted to R. Stora 
for helpful comments on a preliminary draft of this paper.


\end{document}